\documentstyle[12pt]{article}
\begin{document}
\textwidth 16.0cm
\textheight 24.0cm
\topmargin -1.5cm
\oddsidemargin +0.2cm
\evensidemargin -1.0cm

\def\beq{\begin{equation}}
\def\eeq{\end{equation}}
\def\bea{\begin{eqnarray}}
\def\eea{\end{eqnarray}}
\def\bq{\begin{quote}}
\def\eq{\end{quote}}
\def\ve{\vert}

\def\nnb{\nonumber}
\def\ga{\left(}
\def\dr{\right)}
\def\aga{\left\{}
\def\adr{\right\}}

\def\rar{\rightarrow}
\def\nnb{\nonumber}
\def\la{\langle}
\def\ra{\rangle}
\def\nin{\noindent}
\pagestyle{empty}
\begin{flushright}
{CERN-TH.7237/94}\\
%PM 94/xx
\end{flushright}

\vspace*{5mm}
\begin{center}
\section*{
{{
Dominance of the light-quark condensate \\
in the heavy-to-light exclusive decays}}}
\vspace*{1.5cm}
{\bf S. Narison} \\
\vspace{0.3cm}
Theoretical Physics Division, CERN\\
CH - 1211 Geneva 23\\
and\\
Laboratoire de Physique Math\'ematique\\
Universit\'e de Montpellier II\\
Place Eug\`ene Bataillon\\
F-34095 - Montpellier Cedex 05\\
\vspace*{2.0cm}
{\bf Abstract} \\ \end{center}
\vspace*{2mm}
\noindent
Using the QCD {\it
hybrid} (moments-Laplace) sum rule,
 we show $semi$-$analytically$
that, in the limit $M_b \rar \infty$, the $q^2$ and $M_b$ behaviours of
the heavy-to-light exclusive ($\bar B\rar \rho~(\pi)$ semileptonic
as well as the $ B\rar \rho\gamma$ rare)
decay--form factors
 are $universally$ dominated by the contribution
of the soft light-quark condensate rather
than that of the hard perturbative
diagram. The QCD-analytic $q^2$ behaviour of the form factors is
a polynomial in $q^2/M^2_b$, which mimics quite well the usual
pole parametrization, except
in the case of the $A_1^B$ form factor, where there is a significant
deviation
from this polar form.
The $M_b$-dependence of the form factors
expected from HQET
and lattice results is recovered.
We extract with a good accuracy the ratios:
$V^B(0)/A^B_1(0) \simeq A^B_2(0)/A^B_1(0) \simeq 1.11\pm 0.01$,
and
$A^B_1(0)/F^B_1(0) \simeq 1.18 \pm 0.06$; combined with
the ``world average" value of $f^B_+(0)$ or/and $F^B_1(0)$, these ratios
 lead to the decay rates:
$\Gamma_{\bar B\rar \pi e\bar \nu} \simeq (4.3 \pm 0.7)
\times|V_{ub}|^2 \times 10^{12
}$ s$^{-1}$,
$\Gamma_{\bar B\rar \rho e\bar \nu}/
\Gamma_{\bar B\rar \pi e\bar \nu} \simeq .9 \pm 0.2$, and to the ratios
of the $\rho-$polarised rates: $\Gamma_+/\Gamma_- \simeq 0.20 \pm 0.01,
{}~\alpha \equiv 2\Gamma_L/\Gamma_T-1 \simeq -(0.60 \pm 0.01)$.
\vspace*{2.0cm}
\noindent
%\rule[.1in]{15.0cm}{.002in}

%\vspace*{1.5cm}

\begin{flushleft}
CERN-TH.7237/94 \\
%PM 94/xx\\
\today
\end{flushleft}
\vfill\eject
\pagestyle{empty}
%\clearpage\mbox{}\clearpage

\setcounter{page}{1}
\pagestyle{plain}

\section{Introduction}
In previous papers \cite{SN1,SN2}, we have introduced the {\it
hybrid} (moments-Laplace) sum rule (HSR),
which is more appropriate than
the {\it popular} double exponential Laplace (Borel) sum rule (DLSR)
for
studying the
form factors of a heavy-to-light quark transition; indeed,
the {\it hybrid} sum rule has a well-defined
behaviour when the heavy quark mass tends to infinity. In \cite{SN2},
we studied analytically with the HSR
the $M_b$-dependence of the $B\rar K^*
\gamma$ form factor and found that it is dominated by the
light-quark condensate and behaves
like $\sqrt{M_b}$
at $q^2=0$.
 We have also
noticed in \cite{SN1} that the light-quark condensate effect is
important in the numerical evaluation of the $\bar B \rar \rho~(\pi)~
$ semileptonic
form factors, while it has been noticed numerically
in \cite{DOSCH} using the
DLSR that for the $\bar B \rar \rho$ semi-leptonic decays,
the
$q^2$ behaviour of the $A^B_1$ form factor in the time-like region
is
very different from the one expected from
the $standard$ pole representation.
 In this paper, we shall study analytically the
$M_b$-behaviour of the different form factors
for a better understanding of the previous numerical
observations. As a consequence, we shall re-examine
with our analytic expression
the validity of the
$q^2$-dependence obtained numerically
in \cite{DOSCH}, although we shall mainly concentrate
our analysis
in the Euclidian region ($q^2 \le 0$).
There, the QCD calculations
of the three-point function are reliable; also
the lattice
results have more data points.
For this purpose, we shall analyse the
form factors of the $\bar B \rar \pi (\rho)~
$
semileptonic and $B\rar \rho\gamma$ rare
processes defined in a standard way as:
\bea
\la\rho(p')\ve \bar u \gamma_\mu (1-\gamma_5) b \ve B(p)\ra
&=&(M_B+M_\rho)A_1
\epsilon^*_\mu -\frac{A_2}{M_B+M_\rho}\epsilon^*p'(p+p')_\mu \nnb \\
&&+\frac{2V}{M_B+M_\rho} \epsilon_{\mu \nu \rho \sigma}p^\rho p'^\sigma ,
\nnb \\
\la\pi(p')\ve
\bar u\gamma_\mu b\ve B(p)\ra &=& f_+(p+p')_{\mu}+f_-(p-p')_\mu , \nnb \\
<\rho(p')\ve \bar s \sigma_{\mu \nu}\ga
\frac{1+\gamma_5}{2}\dr q^\nu b\ve B(p)> &=&
i\epsilon_{\mu \nu \rho \sigma}\epsilon^{*\nu}p^\rho p'^\sigma
F^{B\rar\rho}_1
\nnb \\
&&+ \aga \epsilon^*_\mu(M^2_B-M^2_{\rho})-\epsilon^*q(p+p')_{\mu}
\adr \frac{F^{B\rar \rho}_1}{2}.
\eea
In the QCD spectral sum rules
(QSSR) evaluation of the form factors, we shall consider the
generic three-point function:
\beq
V(p,p',q) = -\int d^4x \int d^4y\, \mbox{exp}
(ip'x-ipy) \la0\ve TJ_L(x)O(0)J_b(y)\ve
0\ra,
\eeq
whose Lorentz decompositions are analogous to the previous hadronic
amplitudes. Here
$J_L \equiv \bar u\gamma_\mu d
{}~~(J_L\equiv (m_u+m_d) \bar u i\gamma_5 d)$
 is the bilinear quark current having the quantum
numbers of the $\rho~(\pi) $ mesons; $J_b \equiv (M_b +m_d)
\bar d i\gamma_5 b$ is
the quark current associated to the $B$-meson;  $O\equiv \bar b\gamma_\mu
u$ is the charged weak
current for the semileptonic transition, while $O\equiv \bar
b\frac{1}{2}\sigma^{\mu\nu}q_\nu$ is the penguin operator for
the  rare decay.
The vertex function obeys the double dispersion relation:
\beq
V(p^2,p'^2,q^2)= \frac{1}{4\pi^2}\int_{M^2_b}^{\infty}\frac{ds}{s-p^2}
\int_{0}^{\infty}\frac{ ds'}{s'-p'^2} \,
\mbox{Im}\, V(s,s',q^2)+...
\eeq
As already emphasized in \cite{SN2}, we shall
work with the HSR:
\bea
\cal {H}(n,\tau) &\equiv &
\frac{1}{n!}\ga\frac{\partial}{\partial p^2}\dr^n
_{p^2=0}\cal{L}\ga V(p^2,p'^2,q^2) \dr  \nnb \\
&=&
\frac{1}{\pi^2}\int_{M^2_b}^{\infty}\frac{ds}{s^{n+1}}
\int_{0}^{\infty} ds' \, \mbox{exp}(-\tau' s')
\mbox{Im}\, V(s,s',q^2),
\eea
rather than with the DLSR
($\cal{L}$ is the Laplace transform operator). This
sum rule guarantees that
terms of the
type:
\beq
\frac{M^{2l}_b}{\ga M^2_b-p^2 \dr^{k} p'^{2k'}},
\eeq
which appear
in the successive evaluation of the Wilson coefficients of high-dimension
operators, will not spoil the OPE
 for $M_b \rar \infty$ unlike the case of
the double Laplace transform sum rule, which blows up in this
limit for some of its applications in the heavy-to-light transitions.

\nin
In order to come to observables, we
insert intermediate states between the charged weak
and hadronic currents in (2),
while we smear the higher-states effects with
 the discontinuity of the QCD
graphs from a threshold $t_c$ ($t'_c$) for the heavy (light)
mesons. Therefore, we have the sum rule:
\bea
\cal{H}_{res}& \simeq& 2C_L f_B \frac{F(q^2)}{M_B^{2n}}
\mbox{exp} \, (-M^2_L\tau)
\nnb \\
&\simeq&
 \frac{1}{4\pi^2}\int_{M^2_b}^{t_c}\frac{ds}{s^{n+1}}
\int_{0}^{t'_c} ds' \, \mbox{exp}(-\tau s')
\mbox{Im}\, V_{{PT}}(s,s',q^2) + {NPT}.
\eea
$PT~ (NPT)$ refers to perturbative (non--perturbative)
contributions; $C_L \equiv f_P M_P^2$ for light pseudoscalar mesons,
while
$C_L\equiv M_V^2/(2\gamma_V)$ for light vector mesons; $M_L$ is the light
meson mass. The
decay constants are normalized as:
\bea
& &
 (m_q+M_Q)\la 0\ve \bar q (i\gamma_5)Q\ve P\ra= \sqrt 2 M^2_Pf_P \nnb \\
& &\la 0\ve \bar q \gamma_\mu Q\ve V\ra
 =\epsilon^*_\mu \sqrt 2 \frac{ M_V}
{2 \gamma_V^2}.
\eea
$F(q^2)$ is the form factor of interest. For our purpose, we shall
consider the
expression of the decay constant
$f_B$ from moments sum rule at the same order (i.e. to leading order)
\cite{SN3}:
\beq
\frac{2f_B^2}{{(M_B^2)}^{n_2-1}}
\simeq
\frac{3}{8\pi^2}M_b^2 \int_{M^2_b}^{t_c}
\frac{ds}{s^{n_2+1}} \; \frac{(s-M_b^2)^2}{s}
-\frac{<\bar qq>}{M_b^{2n_2-1}}\aga 1-\frac{n_2(n_2+1)}{4} \;
\ga \frac{M_0^2}{M_b^2}\dr\adr .
\eeq
For convenience, we shall work with the
non--relativistic energy parameters $E$ and $\delta M_{(b)}$:
\beq
s \equiv (M_b+E)^2 ~~~~~~~~~\mbox{and}~~~~~~~~~
\delta M_{(b)} \equiv M_B-M_b,
\eeq
where, as we saw in the analysis of the two-point correlator,
the continuum energy $E_c$ is
\cite{SN3}:
\bea
E^D_c &\simeq& (1.08 \pm 0.26)~\mbox{GeV} \nnb \\
E^B_c &\simeq& (1.30 \pm 0.10)~\mbox{GeV} \nnb \\
E^{\infty}_c &\simeq& (1.5 \sim 1.7)~\mbox{GeV}.
\eea
In terms of these continuum energies, and
at large values of $M_b $, the decay constant reads \cite{SN3}:
\bea
f^2_B &\simeq &\frac{1}{\pi^2}
\frac{\ga E^B_c \dr ^3}{M_b}\ga \frac{M_B}{M_b} \dr^{2n_2-1}
\Bigg\{ 1-\frac{3}{2}(n_2+1)\ga
\frac{E^B_c}{M_b}\dr \nnb \\
&&+\frac{3}{5}\ga (2n_2+3)(n_2+1)+
\frac{1}{4} \dr \ga \frac{E^B_c}{M_b} \dr ^2
 -\frac{\pi^2}{2}\frac{\la {\bar qq} \ra}{\ga E^B_c\dr ^3} \ga 1-
\frac{n_2(n_2+1)}{4}
\frac{M_0^2}{M^2_b}\dr
\Bigg\}, \nnb \\
\eea
\section{The \boldmath{$\bar B \rar \rho$} semileptonic decay }
The corresponding
form factors defined in (1)
have been estimated with the HSR \cite{SN1} and the DLSR \cite{SN1},
\cite{DOSCH}.
Instead of taking the average values from the two methods as was done in
\cite{SN1}, we shall only consider the HSR estimates, because of the
drawbacks previously found in the DLSR approach:
\beq
A^B_1(0) \simeq 0.16 - 0.41, \; \; \;
A^B_2(0) \simeq 0.26 - 0.58,  \; \; \;
V^B(0) \simeq  0.28 - 0.61.
\eeq
The errors in these numbers
are large, as  the HSR
has no $n$-stability.
In the following, we
derive  semi-analytic formulae for the form factors.
Using the leading order in $\alpha_s$ QCD results of the
three-point function,
and including the effect of the dimension-5
operators as given in \cite{OVI}, one deduces the sum rule
($q^2 \le 0$):
\beq
A^B_1(q^2) \simeq
-\frac{1}{2}~{\la \bar qq \ra }~
\frac{ \rho_1}{f_B}\ga \frac{M_B}{M_b}\dr^{2n}
\aga 1-\frac{q^2}{M^2_b}
+\delta^{(5)}+\frac{\cal{I}_1}{M^2_b} \adr,
\eeq
with:
\bea
\rho_1 &\equiv&
\ga \frac{\gamma_\rho}{M^2_\rho}\dr \frac{M_b}{(M_B+M_\rho)}
\mbox{exp}(M^2_\rho \tau') \nnb \\
\delta^{(5)}&\equiv& \frac{\tau'M^2_0}{6}\Bigg\{
 n-\frac{1}{\tau'M^2_b}
\ga 1-\frac{3}{4}n-\frac{3}{4}n^2\dr \nnb \\
&&-\frac{q^2}{M^2_b}\ga (n+1)\ga\frac{3}{2}n-1\dr
+2\tau' M^2_b(1+2n)+2(n+1)q^2\tau'\dr \Bigg\}  \nnb \\
%\delta^{(6)}&\equiv& -\frac{1}{M^3_b}
%\ga \frac{8\pi}{81}\dr \frac{\la O_6 \ra}{
%\la \bar qq \ra}
% \aga 4(n+1)-\tau'M^2_b \ga 10-3n+n^2 \dr +\frac{15}{2} (\tau'M^2_b)^2
%-\frac{1}{2}(\tau'M^2_b)^3 \adr  \nnb \\
\eea
where $\cal{I}_1$ is the integral from the perturbative expression
of the spectral function. It is constant for $M_b \rar \infty$. Its
value and behaviour at finite values of $M_b$ and for $q^2=0$
is given in Fig. 1.
At $M_b=4.6$ GeV,
it reads: $\cal{I}_1 \simeq (3.6 \pm 1.2)~\mbox{GeV}^2$ and behaves
to leading order in $1/M_b$ as $t'^2_cE_c/\la \bar qq \ra$, which is
reassuring as it gives a clear meaning of the expansion in (13).
 The other
values of the QCD parameters are \cite{SN4}:
$\la \bar qq \ra =-(189~\mbox{MeV})^3 \ga\log {M_b/\Lambda}\dr^{12/23}$
and $M^2_0 = (0.80\pm 0.01)~ \mbox{GeV}^2$
 from the analysis of the $B,B^*$ sum rules.
%$
% \la O_6\ra \equiv \rho \alpha_s \la \bar qq \ra ^2
% \simeq (3.8 \pm 2.0)\times 10^{-4} \mbox{GeV}^6.$
  The $\rho$-meson coupling is $\gamma_\rho \simeq 2.55$.

\nin
One can deduce from the previous expression that $A^B_1$
is dominated by the light-quark condensate
in the $1/M_b$-expansion counting rule. Moreover,
 the perturbative contribution
is also numerically small at the $b$-mass. The absence of the
 $n$-stability
is explicit from our formula, due to the meson-quark
mass difference entering  the overall factor.
 This effect could be however minimized by using
the expression of $f_B$ in (11) and by imposing that the effects due
to the meson--quark
mass differences from the three- and two-point functions compensate
each other
 to leading order. This is realized by
choosing:
\beq
2n=n_2-\frac{1}{2},
\eeq
which,
fixes $n$ to be about 2,
in view of the fact that the two-point function stabilizes for
$n_2 \simeq$4-5.
In this way, one would obtain the leading-order result in $\alpha_s$:
\beq
A_1^B \simeq  0.3 - 0.6,
\eeq
where we have used the leading-order
value $f^{L.O}_B \simeq 1.24 f_\pi$.
However, although this result is consistent with previous numerical fits
in (12) and in \cite{DOSCH}, we only consider it as an indication of a
consistency rather than a safe estimate because of the previous drawbacks
for the $n-$stability.
One should also keep it in
mind that the values given in (12) correspond to the
value of $f_B \simeq 1.6 f_\pi$, which includes
the radiative corrections of the two-point correlator
and which corresponds to smaller values of $n$. Improvements
 of the result
in (16) need (of course)
an evaluation of the radiative corrections for the
three-point function.
The $q^2$--dependence of $A^B_1$ can be obtained with good accuracy,
without imposing the previous constraint. We obtain the numerical
result in Fig. 2, which is well approximated by the effect from
the light--quark condensate alone:
\beq
R^B_1(q^2) \equiv \frac{A_1^B(q^2)}{A_1(0)}
\simeq 1-\frac{q^2}{M^2_b}.
\eeq
Performing an analytic
continuation of this result in the time-like region, we reproduce the
numerical result from the DLSR \cite{DOSCH}(see Fig. 2),
which indicates that the
result is independent of the form
of the sum rule used, while in the time-like region the
perturbative contribution still remains a small correction of
the light-quark condensate one.
 This result is clearly in contradiction with
the $standard$ pole-dominance parametrization, as, indeed,
the form factor decreases for increasing $q^2$-values.
 A test of this result
needs improved lattice measurements over the ones available in
\cite{SACH}.
 From the previous
expressions, and using the fact that $f_B$ behaves as $1/\sqrt{M_b}$,
one can also predict the $M_b$-behaviour of the form factor
at $q^2_{max} \simeq M^2_b+2M_\rho M_b$:
\beq
A^B_1(q^2_{max}) \sim \frac{1}{\sqrt{M_b}},
\eeq
in accordance with the expectations from HQET \cite{HQET} and
the
lattice results \cite{SACH}.
The analysis of the $V^B$ and $A^B_2$ form factors will be done in the
same way. Here, one can realize that the inclusion of the higher
dimension-5 and -6
 condensates tends to destabilize the results, although
these still remain small corrections to the leading-order results.
Then, neglecting these destabilizing terms, one has:
\bea
V^B(q^2) \simeq
-\frac{1}{2}~\la\bar qq\ra~\frac{\rho_V}{f_B}~
\ga \frac{M_B}{M_b} \dr^{2n}
  \aga 1  + \frac{\cal{I}_V}{M^2_b}+... \adr \nnb \\
A^B_2(q^2) \simeq
-\frac{1}{2}~\la\bar qq\ra~ \frac{\rho_{2}}{f_B}~
\ga \frac{M_B}{M_b} \dr^{2n}
  \aga 1  + \frac{\cal{I}_2}{M^2_b}+... \adr
\eea
with:
\bea
\rho_V &\equiv&
\ga \frac{\gamma_\rho}{M^2_\rho}\dr \frac{M_b(M_B+M_\rho)}{M^2_B}
\mbox{exp}(M^2_\rho \tau')  \nnb \\
\rho_2 &\equiv&
\ga \frac{\gamma_\rho}{M^2_\rho}\dr \frac{(M_B+M_\rho)}{M_b}
\mbox{exp}(M^2_\rho \tau').
\eea
$\cal{I}_{V,2}$ are integrals from the perturbative spectral functions,
which also behave like $\cal{I}_1$ to leading order in $1/M_b$.
They are given in Fig. 1 for $q^2=0$ and for different values of $M_b$.
As expected, they are constant when $M_b \rar \infty$, although, as in
the previous case, the asymptotic limit is reached very slowly.
Here, the $n$-stability of the analysis
is also destroyed by the overall
$(M_B/M_b)^{2n}$ factor, which hopefully disappears when we work with
the ratios of form factors. We show in Fig. 2 the $q^2$-dependence
of the normalized
$V^B$ and $A^B_2$, which is very weak since the dominant
light-quark condensate contribution has no $q^2$-dependence.
The small increase with $q^2$ is due to the $q^2$-dependence
of the small and non-leading contribution from the
perturbative graph. Lattice points in the Euclidian
$q^2$-region \cite{SACH} agree with our results.
An analytic continuation of our
results at time-like $q^2$ agrees qualitatively
with the one in \cite{DOSCH}. The numerical difference in this
region is due to the relative increase of the perturbative contribution
in the time-like region due to the effect of the
additional non-Landau-type singularities.
However, this effect does not influence
the $M_b$ behaviour of the form factors
at $q^2_{max}$, which
can be safely
obtained from
the leading-order expression given by the light-quark condensate. One
can deduce:
\beq
V^B(q^2_{max}) \sim  \sqrt{M_b},~~~~~~~
A^B_2(q^2_{max}) \sim  \sqrt{M_b}.
\eeq
This result is in agreement with HQET and lattice data
points.
Finally, we can also extract the ratios of form factors.
At the $\tau'$-maxima and at the $n$-maxima or inflexion point, we
obtain from Fig. 3:
\beq
r_2 \equiv \frac{A^B_2(0)}{A^B_1(0)} \simeq
r_V \equiv \frac{V^B(0)}{A^B_1(0)} \simeq  1.11 \pm 0.01,
\eeq
where the accuracy is obviously due to the cancellation of systematics
in the ratios. This result is again consistent with the lattice results
\cite{SACH}, but more accurate.
\section{ The \boldmath{$\bar B \rar \pi~$}semileptonic decay}
The relevant form factor defined in (1) has been
numerically estimated within the HSR with the value \cite{SN1}:
\beq
f^B_+(0) \simeq 0.20 \pm 0.05,
\eeq
%(the average from HSR and DLSR is $0.23 \pm 0.02$ which is
%consistent with other predictions \cite{DOSCH}),
where the contribution of the $\pi'$(1.3) meson has been included
for improving the sum rule variable stability of the result. In
this paper, we propose to explain the meaning of this numerical
result from an analytic expression of the sum rule. Using the QCD
expression given in \cite{OVI}, we obtain for a pseudoscalar current
describing the pion:
\beq
f^B_+(q^2) \simeq -\frac{(m_u+m_d)\la \bar qq \ra}{4f_\pi m^2_\pi}
\frac{1}{f_B} \ga\frac{M_B}{M_b}\dr^{2n}
\aga 1+\delta^{(5)}+\frac{\cal{I}_\pi}{M^2_b} \adr ,
\eeq
where $\cal{I}_\pi$ is the spectral integral coming from the
perturbative graph. Its value at $q^2=0$ for different values of $M_b$
is shown in Fig. 1. It indicates that at $M_b=4.6$ GeV, the perturbative
contribution, although large, still
 remains a correction compared with the light-quark
condensate term; $\delta^{(5)}$ is the correction due to the
dimension-5 condensate and reads:
\beq
\delta^{(5)} \simeq -\frac{\tau'M^2_0}{6} \aga 2n+
\frac{\tau'^{-1}}{4M^2_b}(n+1) \ga \frac{3}{2}n-1\dr \adr .
\eeq
One can use the well-known PCAC relation
\beq
(m_u+m_d) \la \bar qq \ra = -m^2_\pi f^2_\pi, ~~~~~~~~~~~~~f_\pi=93.3
{}~\mbox{MeV}
\eeq
into the previous sum rule in order to express $f^B_+$ in terms
of the meson couplings. Unlike the case of the $B\rar \rho$
form factors where the scale dependence is contained in
$\la \bar qq \ra$,
$f^B_+$ is manifestly renormalization-group-invariant. It should
be noted, as in the case of the sum rule
determination of the $\omega\rho\pi$ coupling \cite{SN4}, that
the $f_\pi$-dependence appears indirectly via (26) in a correlator
evaluated in the deep Euclidian region, while the pion is off shell,
which is quite different
  from soft-pion techniques with an on-shell Goldstone boson.
One can also
deduce from (24) that for large $M_b$, $f^B_+$ behaves like
$\sqrt{M_b}$. In this limit the
$q^2$-dependence is rather weak, as it
comes only from the non-leading $1/M_b$ contributions; we therefore
have, to a good accuracy:
\beq
f^B_+(q^2_{max}) \simeq f^B_+(0) \sim \sqrt{M_b}.
\eeq
As in the previous case,
 the slight difference
between the $q^2$-behaviour
in the time-like region and the one from that
obtained in \cite{DOSCH},
at a finite value of $M_b$(=4.6 GeV),
is only due
to a numerical enhancement caused by the non-Landau singularities
of the perturbative contribution
in this region, but does not disturb the $M_b$-behaviour of the form
factor. Finally, we extract the ratio of the form factor:
\beq
r_\pi \equiv \frac{A^B_1(0)}{f^B_+(0)}.
\eeq
Unfortunately, we do not have stabilities, as the stability points
are different for each form factor, which is mainly due
to the huge mass-difference between the $\rho$ and $\pi$ mesons.
\section{{\boldmath{The $B \rar \rho \gamma$}} rare decay}
We can use the previous results into the HQET \cite{HQET}
relation among the different form factors of the
rare
$B \rar \rho \gamma$ decay
($F_1^{B}\equiv
F_1^{B \rar \rho}$)
and the semileptonic ones. This relation reads around $q^2_{max}$:
\beq
F_1^{B }(q^2)
= \frac{q^2+M^2_B-M^2_{\rho}}{2M_B}
\frac{V^B(q^2)}{M_B+M_{\rho}}+\frac{M_B+M_{\rho}}{2M_B}A^B_1(q^2),
\eeq
from which we deduce:
\beq
F_1^{B}(q^2_{max}) \sim \sqrt{M_b}.
\eeq
However,
we can also
study, directly from the sum rule, the $q^2$-dependence of
$F_1^{B }$. Using the fact
that the corresponding sum rule is also
dominated by the light-quark condensate for $M_b \rar \infty$
\cite{SN2},
an evaluation of this
contribution, at $q^2 \not= 0$,
shows that the light-quark condensate effect has no
$q^2$-dependence to leading order. Then, we can deduce, to a good
accuracy:
\beq
F_1^B(q^2_{max}) \simeq  F_1^B(0) \sim \sqrt{M_b}.
\eeq
Let us now come back to the parametrization of the form factor at
$q^2=0$. We have given in \cite{SN2} an expanded interpolating
formula that involves  $1/M_b$ and $1/M^2_b$
corrections due to the meson-quark mass difference, to $f_B$ and to
higher-dimension condensates.
 Here, we present
a slightly modified expression, which is:
\beq
F_1^B(0) \simeq
-\frac{1}{2}~\la\bar qq\ra~\frac{\rho_\gamma}{f_B}~
\ga \frac{M_B}{M_b} \dr^{2n}
  \aga 1  + \frac{\cal{I}_\gamma}{M^2_b}+... \adr,
\eeq
with:
\bea
\rho_\gamma &\equiv&
\ga \frac{\gamma_\rho}{M^2_\rho}\dr
\mbox{exp}(M^2_\rho \tau'), \nnb \\
\cal{I}_\gamma &\simeq& (20\pm 4) \mbox{ GeV}^2
{}~~~~~ \mbox{for}~ M_b \ge 4.6~ \mbox{GeV},
\eea
where  we have neglected the effects of higher-dimension
condensates;
$\cal{I}_\gamma$
is the perturbative spectral integral. One should notice that unlike
the
other spectral integrals in Fig. 1, $\cal{I}_\gamma$ reaches quickly the
asymptotic limit when $M_b \rar \infty$.
Using the estimated value of $F^B_1(0)$ in \cite{SN2},
we can have, in units of GeV:
\beq
F^B_1= \frac{1.6\times 10^{-2}}{f_B}\ga 1+\frac{20\pm 4}{M^2_b} \dr,
\eeq
which leads of course to the same formula at large $M_b$
as in \cite{SN2}. However,  due to the large coefficient of the
perturbative contribution,
it indicates
that an extrapolation of the result obtained
at low values of $M_c$ is quite
dangerous, as it may lead to a wrong  $M_b$-behaviour of the form
factor at large mass. One should notice that (34) and the one in
\cite{SN2} lead to the same numerical value of
$F^D_1(0)$. Proceeding as for the former cases, we can also extract
the ratio:
\beq
r_\gamma \equiv \frac{A_1^B(0)}{F_1^B(0)} \simeq 1.18\pm 0.06,
\eeq
from the analysis of the $\tau'$- and $n$-stability shown in Fig. 3.

\section{Values of the \boldmath{$B$}-form factors}
The safest prediction of the absolute value of the form factors
available at present, where different versions
of the sum rules and lattice calculations have a consensus, is the
one for $f^B_+(0)$:
\bea
f^B_+(0) \simeq & &0.26 \pm 0.12 \pm 0.04~~~~~~\mbox{Lattice}~
 \cite{SACH}\nnb \\
                & &0.26 \pm 0.03 ~~~~~~~~~~~~~~~
                \mbox{DLSR}   ~\cite{DOSCH}
(\mbox{see also} \cite{OVCH})                 \nnb \\
                & &0.23 \pm 0.02 ~~~~~~~~~~~~~~~
                 \mbox{HSR+DLSR}~\cite{SN1} \nnb \\
                & &0.27 \pm 0.03 ~~~~~~~~~~~~~~~
                 \mbox{Light-cone}~\cite{RUCKL},\nnb \\
\eea
from which one can deduce the ``world average":
\beq
f^B_+(0) \simeq  0.25 \pm 0.02 .
\eeq
For estimating $A^B_1(0)$,
one can use the present most reliable estimate
of $F^B_1$
\cite{SN2}, \cite{ALI}:
\beq
F^B_1(0) \simeq 0.27 \pm 0.03,
\eeq
where we have used the strength of the $SU(3)$-breakings obtained
in \cite{SN2}, in order to convert the result for $B\rar K^*\gamma$
of \cite{ALI} into the $B\rar \rho\gamma$ of interest here. Then,we
deduce:
\beq
A_1^B(0) ~ \simeq ~ 0.32 \pm 0.02,~~~
\eeq
which is consistent with a direct estimate \cite{SN1,ALI}, but the
result is again more accurate.
\section{\boldmath{$B$}-semileptonic-decay rates}
We are now in a good position to predict the different decay rates.
In so doing, we shall use the pole parametrization, except for the
$A^B_1$ form factor. For the $B\rar \pi$, we shall use the experimental
value 5.32 GeV of the $B^*$ mass. For the $B\rar \rho$, we shall
use the fitted value ($6.6\pm 0.6$) GeV \cite{DOSCH}
for the pole mass associated to
$A^B_2$ and $V^B$.
For $A^B_1$, we use the linear
form suggested by (13), with an effective mass of ($5.3\pm 0.7$) GeV,
which
we have adjusted from the numerical behaviour given in \cite{DOSCH}
(we have not tried to reproduce the change of the behaviours for
 $t\simeq
(0.76-0.95)M^2_b$ obtained in \cite{DOSCH}, which is a minor effect).
Using the standard definitions and notations,
we obtain:
\beq
\Gamma_{\bar B\rar \pi e\bar \nu} \simeq (3.6 \pm 0.6)\times
 |V_{ub}|^2\times 10^{12}~\mbox{s}^{-1}
\eeq
We also obtain the following ratios:
\beq
\frac{\Gamma_{\bar B\rar \rho e\bar \nu}}
{\Gamma_{\bar B\rar \pi e\bar \nu}} \simeq 0.9 \pm 0.2,~~~~~
\frac{\Gamma_+}{\Gamma_-} \simeq 0.20 \pm 0.01,~~~~~
\alpha \equiv 2\frac{\Gamma_L}{\Gamma_T}-1
 \simeq -(0.60 \pm 0.01).
\eeq
Thanks to a better control of the ratios of form factors, the ratio
of the $\bar B$
decays into $\pi$ over the $\rho$ can be predicted, to a good
accuracy. It becomes compatible with the prediction obtained by
only retaining the contribution of the vector component of the
form factors. Our predictions are compatible with the ones in
\cite{DOSCH} except
for $\Gamma_+/\Gamma_-$, where the one in \cite{DOSCH} is about one
order of magnitude smaller. The difference of two of
these three quantities
with ones in \cite{SN1}
(the large branching ratio into  $\rho$ over $\pi$ and the
positive value of the asymmetry $\alpha$ in \cite{SN1} and in most
other pole dominance models for $A^B_1$)
is mainly due to the different $q^2-$behaviour
of $A^B_1$ used here.
\section{Conclusions}
We have studied, using the QCD $hybrid$ sum rule, the $M_b$- and
$q^2$-behaviours of the heavy-to-light transition form factors. We
find that these quantities are dominated in a $universal$ way
by the light-quark condensate
contribution.

\nin
The $M_b$-dependence obtained here
is in perfect agreement with the HQET and lattice
expectations.

\nin
The $q^2$-dependence of the $A^B_1$ form factor,
which is mainly due to the one from the light-quark condensate
contribution, is in
clear contradiction with the one expected from a pole parametrization.
The other form factors can mimic $numerically$ this pole parametrization.
Our QCD-analytic $q^2$-behaviours confirm the previous numerical
results given in \cite{DOSCH}.

\nin
We have also shown that it can be incorrect to derive the
 $M_b$-behaviour of the form factors at $q^2=0$ by combining the HQET
 result at $q^2_{max}$ with the pole parametrization.

\nin
We have also shown that the unusual $q^2-$behaviour of the
$A^B_1$ form factor affects strongly the  branching ratio of $B \rar \rho
$ over $B\rar \pi$ and the $\rho$-polarisation
parameter $\alpha$. A measurement
of these quantities complemented by the one of the
$q^2-$behaviour of the form factor should provide a good test of the
sum rules approach.

\nin
 We want also to stress that the extrapolation of the results
obtained in this paper to the case of the $D$-meson would
 be too audacious:
 the uses of the HSR in that case cannot be $rigorously$
justified since the value of the $c$-quark mass is smaller,
although it may lead to acceptable phenomenological results.
We are investigating this point at present.
%\vfill\eject
\section*{Acknowledgements}
I thank
Olivier P\`{e}ne and Chris Sachrajda for discussions of the lattice
results.
\noindent
\section*{Figure captions}

\nin
{\bf Fig. 1}~~~~ $M_b$-dependence of the perturbative spectral integrals
at $q^2$ = 0.
\vspace*{0.5cm}

\nin
{\bf Fig. 2}~~~~$q^2$-behaviour of the normalized form factors:
 $R_1 \equiv A_1^B(q^2)/A_1^B(0)$,
 $R_2 \equiv A_2^B(q^2)/A_2^B(0)$,
 $R_V \equiv V^B(q^2)/V^B(0)$ and
 $R_\pi \equiv f_+^B(q^2)/f_+^B(0)$. The squared points in the timelike
region are from \cite{DOSCH}.
\vspace*{0.5cm}

\nin
{\bf Fig. 3}~~~~$\tau'$- and $n$-dependences of the ratios of form
factors
at $q^2$ = 0:
 $r_2 \equiv A_2^B(0)/A_1^B(0)$,
 $r_V \equiv V^B(0)/A_1^B(0)$ and
 %$r_\pi \equiv A_1^B(0)/f_+^B(0)$ and
 $r_\gamma \equiv A_1^B(0)/F_1^B(0)$.
\vfill\eject


\noindent
\begin{thebibliography}{999}

\bibitem{SN1}S. Narison,
{\it Phys. Lett.} {\bf B283} (1992) 384;
{\it Z. Phys.} {\bf C55} (1992) 55.
\bibitem{SN2}S. Narison,
 ({\it Phys. Lett.}{\bf B327} (1994) 354.
\bibitem{DOSCH}
P. Ball,
{\it Phys. Rev.} {\bf D48} (1993)
 3190; \\
P. Ball, V.M. Braun and H.G. Dosch,
{\it Phys. Rev.} {\bf D44} (1991)
 3567;
{\it Phys. Lett.} {\bf B273} (1991) 316;
\bibitem{SN3}S. Narison,
{\it Phys. Lett.} {\bf B198} (1987) 104;
{\bf B308} (1993) 365
and {\it  Talk given at the Third
$\tau$Cf Workshop, 1-6 June 1993, Marbella, Spain}, CERN preprint
 TH-7042/93 (1993)
and references therein;\\
%{\it Z. Phys.} {\bf C55} (1992) 55;
S. Narison and K. Zalewski,
{\it Phys. Lett.}
{\bf B320} (1994) 369;
\bibitem{OVI}A. Ovchinnikov and V.A. Slobodenyuk,
{\it Z. Phys.} {\bf C44} (1989) 433; \\
V.N. Baier and A.G. Grozin,
{\it Z. Phys.} {\bf C47} (1990) 669.
\bibitem{SN4}S. Narison,
 {\it Lecture Notes in Physics, Vol. 26,
QCD Spectral Sum Rules} (World Scientific, Singapore, 1989)
and references therein;
\bibitem{SACH}ELC: A. Abada et al.,
LPTHE Orsay-93/20 (1993).
\bibitem{HQET}N. Isgur and M.B. Wise,
{\it Phys. Rev.} {\bf D42} (1990)
 2388; \\
G. Burdman and J.F. Donoghue,
 {\it Phys. Lett.} {\bf B270} (1991) 55.
%\bibitem{NEU}G. Burdman et al. SLAC-PUB-6345 (1993).
\bibitem{OVCH}A.A. Ovchinnikov,
 {\it Phys. Lett.} {\bf B229} (1989) 127.
\bibitem{RUCKL}V.M. Belyaev, A. Khodjamirian and R. R\"uckl,
{\it Z. Phys.} {\bf C60} (1993) 349.
%V.L. Chernyak and I.R.
%Zhitnitski,
% {\it Nucl. Phys.} {\bf B345} (1990) 137.
\bibitem{ALI}A. Ali, V.M. Braun and H. Simma, CERN preprint TH-7118/93
(1993).
%\bibitem{MART} private communication from G. Martinelli and
%C. Sachrajda.
%\bibitem{COL}P. Colangelo, C.A. Dominguez, G. Nardulli and N. Paver,
% {\it Phys. Lett.} {\bf B317} (1993) 183;
\end{thebibliography}
\end{document}